\documentclass[prl,twocolumn,groupedaddress]{revtex4-1}

\usepackage[T1]{fontenc}
\usepackage[utf8]{inputenc}
\usepackage[english]{babel}
\usepackage{amsmath,amssymb,graphicx,microtype,siunitx}

\begin{document}

\title{Entropy and disorder enable charge separation in organic solar cells}

\author{Samantha N. Hood}
\affiliation{Centre for Engineered Quantum Systems, Centre for Organic Photonics and Electronics and School of Mathematics and Physics, The University of Queensland, Brisbane QLD 4072, Australia}

\author{Ivan Kassal}
\email{i.kassal@uq.edu.au}
\affiliation{Centre for Engineered Quantum Systems, Centre for Organic Photonics and Electronics and School of Mathematics and Physics, The University of Queensland, Brisbane QLD 4072, Australia}

\begin{abstract}
Although organic heterojunctions can separate charges with near-unity efficiency and on a sub-picosecond timescale, the full details of the charge-separation process remain unclear. In typical models, the Coulomb binding between the electron and the hole can exceed the thermal energy $k_\mathrm{B}T$ by an order of magnitude, suggesting that it is impossible for the charges to separate before recombining. Here, we consider the entropic contribution to charge separation in the presence of disorder and find that even modest amounts of disorder have a decisive effect, reducing the charge-separation barrier to about $k_\mathrm{B}T$ or eliminating it altogether. Therefore, the charges are usually not thermodynamically bound at all and could separate spontaneously if the kinetics otherwise allowed it. Our conclusion holds despite the worst-case assumption of localised, thermalised carriers, and is only strengthened if mechanisms like delocalisation or `hot' states are also present. 

\end{abstract}

\maketitle

\begin{figure}[b]
\centering
\includegraphics[width=0.9\linewidth]{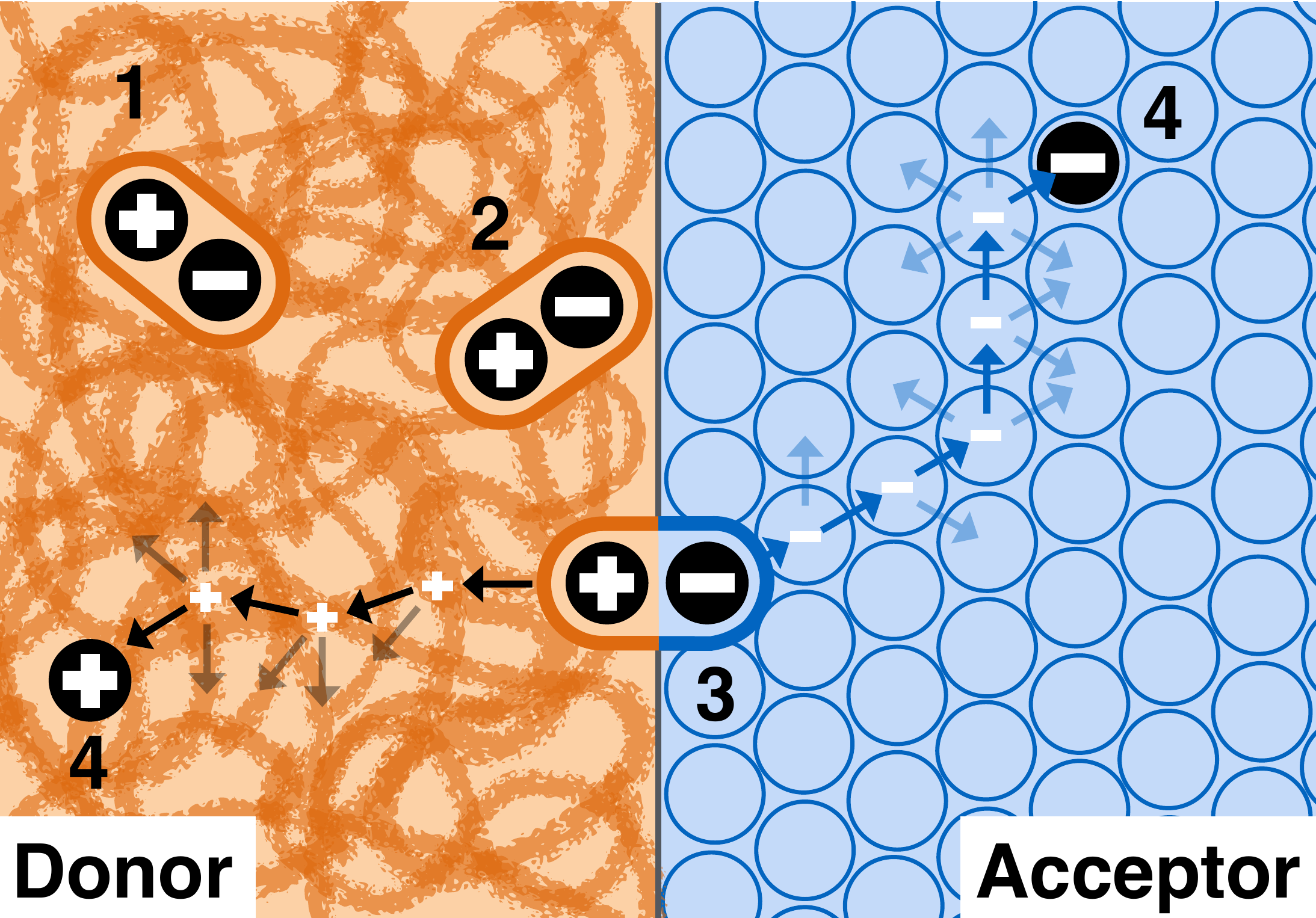}
\caption{Charge separation in organic solar cells. Organic solar cells typically contain two materials, a donor (usually a polymer) and an acceptor (usually a fullerene). After an exciton (1) is formed by the absorption of a photon, it diffuses to the donor-acceptor interface (2). Charge transfer across the interface forms a charge-transfer state (3), whose dissociation into free charges (4) is responsible for charge generation. Arrows: as the charges move away from the interface, they have access to more sites, increasing the entropy.} 
\label{fig:chargetransfer}
\end{figure}

Although organic solar cells (OSCs) have the potential to become low-cost renewable-energy sources, the precise mechanisms of how they convert photoexcitations into free charge carriers are not completely understood.  The efficiency of OSCs depends on the separation of charge-transfer (CT) states formed at donor-acceptor interfaces (Fig.~\ref{fig:chargetransfer}), where the Coulomb binding energy between the hole and the electron is usually assumed to be
\begin{equation}
\label{eq:coulomb}
U(r) = \frac{e^2}{4\pi\varepsilon_0\varepsilon_r r},
\end{equation}
where $r$ is the distance between them, $e$ is the elementary charge, $\varepsilon_0$ is the permittivity of vacuum, and $\varepsilon_r$ is the dielectric constant. Organic semiconductors typically have low dielectric constants, $\varepsilon_r \approx 2 - 4$, meaning that a CT state with a typically assumed nearest-neighbour electron-hole separation of $r = \SI{1}{nm}$ experiences a Coulomb binding of about $\SI{500}{meV}$~\cite{Clarke2010}. Because this barrier is much greater than the available thermal energy ($k_\mathrm{B}T = \SI{25}{meV}$), one might expect that the charges could never separate, or at least not within the lifetime of the CT state. Nevertheless, CT-state dissociation approaches 100\% efficiency in some OSCs~\cite{Park2009}; explaining this has been a central question in the field.

Efficient charge separation can be reproduced using several kinetic models, either kinetic Monte Carlo simulations~\cite{Offermans2005,VanEersel2012,Zimmerman2012,Burke2014} or analytical generalisations of Onsager's theory~\cite{Onsager1934,Onsager1938,Sano1979,Braun1984,Wojcik2009,Hilczer2010,Wojcik2010}. However, it is often difficult to distill the fundamental physics from intricate simulations. In the present case, the simulations have not set aside the widespread view that additional mechanisms are needed to explain charge-separation in OSCs~\cite{Few2014,Gao2014}. Among the proposed mechanisms, charge delocalisation could decrease the Coulomb binding~\cite{Deibel2009, Baranovskii2012, Bernardo2014, Tamura2013, Bakulin2012, Bittner:2014ea, Gagorik2015}, while the excess energy of the initially ``hot'' CT state could help the charges overcome their attraction~\cite{Tamura2013, Bakulin2012, Dimitrov2012, Bassler2015}. 

A related conceptual difficulty is distinguishing bound charges from free ones when the binding potential (such as Eq.~\ref{eq:coulomb}) monotonically increases~\cite{Zaccone2012}. It is conventionally assumed the charges are free if $U(r)<k_\mathrm{B}T$, i.e., if their separation exceeds the Bjerrum length (or Coulomb capture radius) $r_c = e^2/4\pi\varepsilon_0\varepsilon_r k_\mathrm{B}T$. As Onsager already pointed out, this cut-off is ``somewhat arbitrary''~\cite{Onsager1934}. As Onsager already pointed out, this cut-off is ``somewhat arbitrary''~\cite{Onsager1934}. We show that $r_c$ is not the separation at which charges spontaneously dissociate in an OSC and we redefine the capture length to avoid the arbitrariness of this cut-off.

Here, we show that a combination of entropy and energetic disorder suffice to reduce the barrier to CT-state dissociation to make it easily surmountable or entirely absent, even with the worst-case assumption of localised and thermalised charges. Our result can be seen as a simple, thermodynamic explanation of the success of previous kinetic models. We emphasise that, in the absence of a large barrier, purely thermodynamic considerations cannot predict rates $k$, which would be governed by the pre-exponential factors in the Arrhenius equation, $k \propto\exp{\left(-\Delta E/k_\mathrm{B}T\right)}$. Given this expression for the separation rate along with the large energetic barrier $\Delta E$, one might expect the charges never to separate.

Two groups have previously argued that entropy can reduce the height of the potential barrier~\cite{Clarke2010, Gregg2011} because increasing the electron-hole separation $r$ increases the number of ways $\Omega(r)$ for the charges to be arranged, thus decreasing the free energy. Clarke and Durrant~\cite{Clarke2010} showed that the entropic contribution to charge separation is important because it has a similar magnitude to the Coulomb binding energy. They modelled $\Omega(r)$ as scaling as $r^3$ due to their assumption that the electron can occupy sites within a hemisphere centred on the stationary hole. Gregg~\cite{Gregg2011} also considered the entropy of charge separation, but he modelled an exciton in a single material. He recognised that $\Omega(r)$ should scale not with the volume but with the surface area, because it counts the accessible states when the electron and hole are restricted to be $r$ apart. For a given $r$, the electron can occupy sites on a sphere centred at the hole, giving $\Omega(r) = 4\pi (r/a)^2$, where $a$ is the lattice constant. At a planar interface, the number of states on a hemisphere would be $\Omega(r) = 2\pi (r/a)^2$. Subsequently, observations that charges on an interface between an organic semiconductor and vacuum can escape the Coulomb barrier have been attributed to an increase in entropy~\cite{Monahan2015}. 

Here, we address two limitations of the previous models that led them to underestimate the magnitude of entropic effects and retain the conclusion that additional mechanisms might be necessary to explain charge separation. First, we remove the assumed inability of the hole to leave the interface and, second, we add disorder. 

\begin{figure}[t]
\centering
\includegraphics[width=0.9\linewidth]{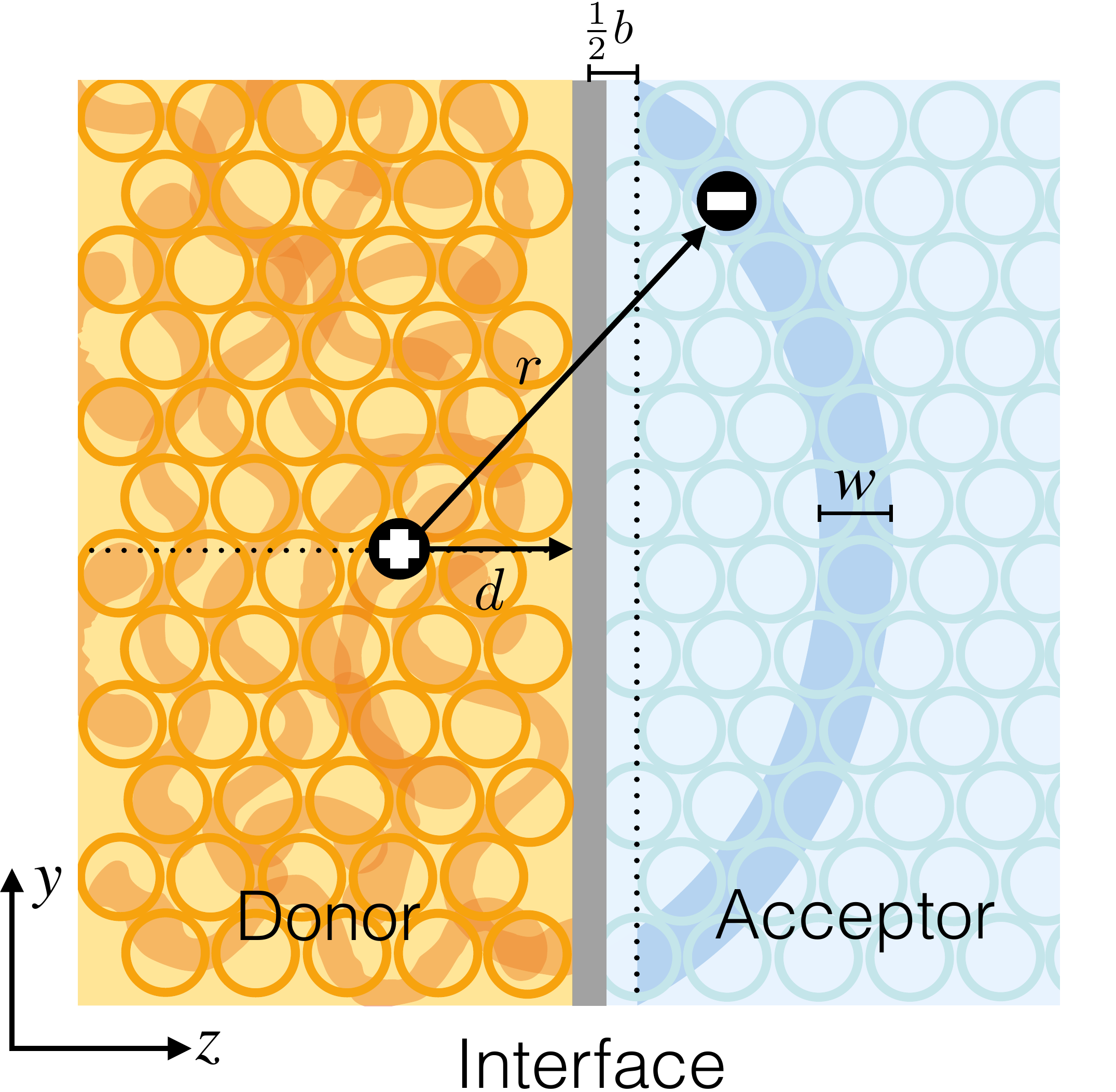}
\caption{Counting the states. The electron and the hole can be at a distance $r$ apart in many different ways, all of which need to be counted to determine the entropy of charge separation. At each distance $d$ between the hole and the interface, the electron can be in a spherical-cap shell of radius $r$ and width $w$ (shown in blue). The number of possible electron-hole configurations is the sum of the volumes of all of the resulting spherical cap shells as $d$ varies from $b/2$ to $r-b/2$.}
\label{fig:Sphericalcap}
\end{figure}

We model the organic semiconductors as a hexagonal close-packed lattice with a site density of $\rho = \SI{1}{nm}^{-3}$~\cite{Clarke2010, Gregg2011}, giving a lattice constant of $a =(\sqrt{2}/\rho)^{1/3} = \SI{1.12}{nm}$. Both the donor and the acceptor have $\varepsilon_r=3.5$ and the temperature is $T=\SI{300}{K}$. The electron and the hole occupy sites on the lattice and start as a CT state at the interface, with the hole in the donor and the electron in the acceptor. The nearest-neighbour distance across the interface (which is in the $xy$ plane) is typically $b = \SI{1}{nm}$~\cite{Clarke2010}, as in Fig.~\ref{fig:Sphericalcap}. 

We consider the energetically ordered case before adding disorder. Because of translational symmetry along $x$ and $y$, the CT state can be assumed without loss of generality to initially be at the origin. Doing so eliminates two degrees of freedom, so it is necessary to eliminate two degrees of freedom from the separated charges as well. We do so by fixing the $x$ and $y$ coordinates of the hole, allowing it to only move along $z$, as shown in Fig.~\ref{fig:Sphericalcap}. Alternatively, the $x$ and $y$ coordinates of the centre of mass of the separated charges could be fixed. 

For the charges to separate, either the electron or the hole (or both) can leave the interface, meaning that allowing only the electron to leave undercounts the number of accessible states (this is so even if the electron is more mobile than the hole, since entropy is a thermodynamic quantity and not a kinetic one). Because all the site energies are the same at a given value of $r$, the ordered case is easily discussed in the microcanonical ensemble, as was done in refs.~\cite{Clarke2010, Gregg2011}. For entropy to be properly defined, it should contain all contributions where the charges are $r$ apart, including when the hole is at different depths, $d$, into the donor. For a given $d$, the electron can occupy sites within a spherical-cap shell centred on the hole and having the desired radius $r$. The number of possible sites for the electron is proportional to the volume of the spherical-cap shell: $n(r,d,b) = 2\pi r (r - d - b/2)w\rho$, where we assume the shell has width $w = \SI{1}{nm}$. The total number of accessible arrangements $\Omega(r)$ is the sum of contributions from all the spherical caps as $d$ varies from $b/2$ (hole adjacent to the interface) to $r-b/2$ (electron adjacent to the interface), i.e.,
\begin{equation}
\Omega(r)  = \sum_{d = \frac{b}{2}}^{r-\frac{b}{2}} n(r,d,b)  = \pi r(r-b)^2\rho, \label{eq:omega}
\end{equation}
where the sum goes in steps of $w$. Finally, with entropy $\Delta S(r)=k_\mathrm{B} \ln \Omega(r)$, the free energy is 
\begin{equation}
\Delta G_\mathrm{ordered}(r)= U(r) - k_\mathrm{B}T\ln \Omega(r). \label{eq:DeltaG}
\end{equation}

If the hole's position is assumed to be fixed~\cite{Gregg2011}, $\Omega(r)$ scales as $r^2$, whereas in the present model it scales as $r^3$ due to the additional degree of freedom of the hole. Because the entropy depends logarithmically on $\Omega(r)$, only the scaling matters; prefactors cancel when computing entropy differences, meaning that it is not important to know $\rho$ and $w$ precisely. The same reasoning implies that a planar interface is a reasonable approximation, since changing the curvature of the interface would mostly affect the prefactor, as opposed to the scaling.

\begin{figure}[t]
\centering
\includegraphics[width=0.9\linewidth]{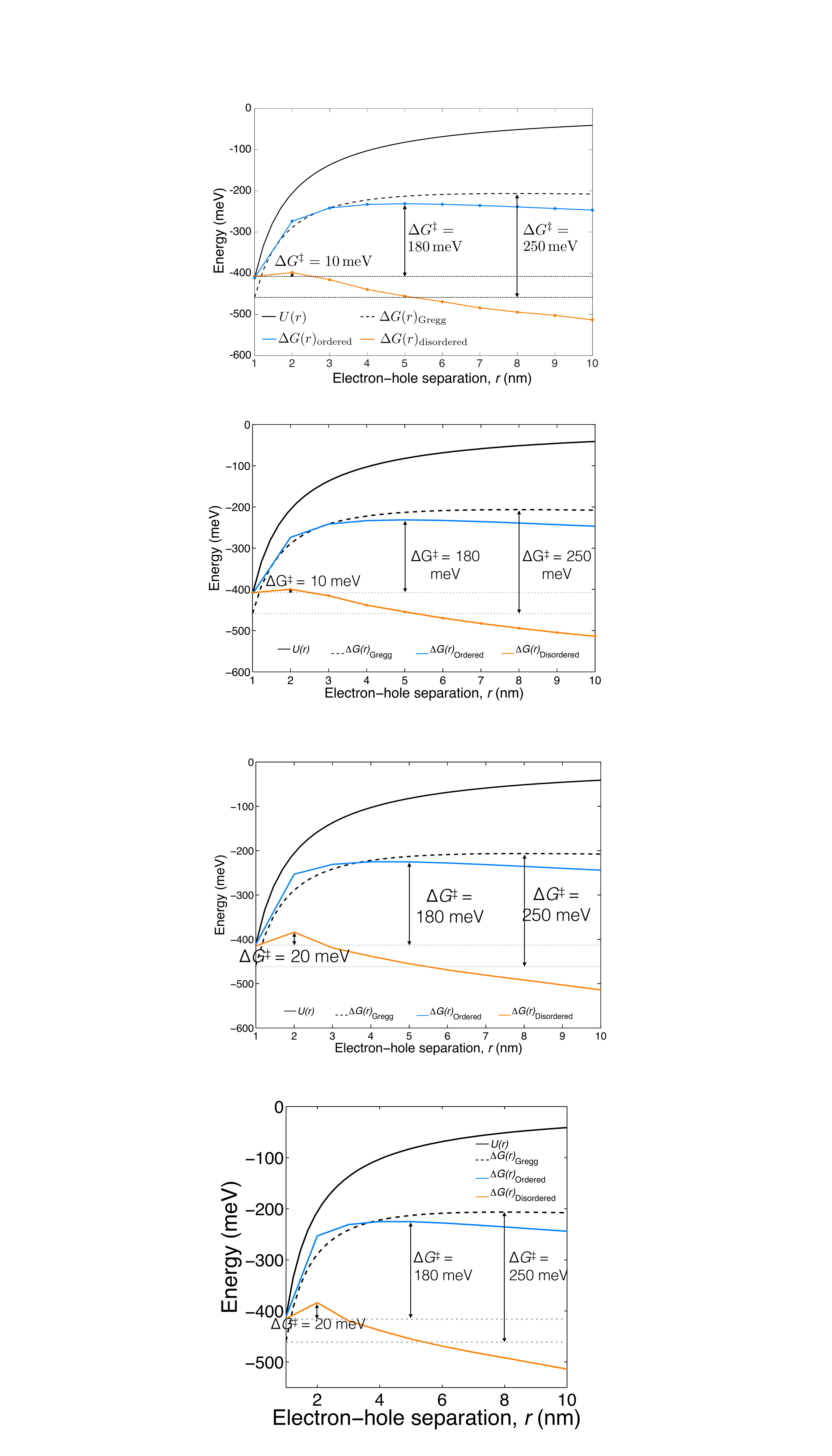}
\caption{The Coulomb potential $U(r)$ and the free energy of the charge-transfer state as a function of the electron-hole separation $r$. $\Delta G_{\mathrm{Gregg}}$ is the free energy in an energetically ordered model with an immobile hole (proposed by Clarke and Durrant~\cite{Clarke2010}, with a correction due to Gregg~\cite{Gregg2011}). $\Delta G_{\mathrm{ordered}}$ assumes both the electron and the hole are able to move, while $\Delta G_{\mathrm{disordered}}$ also includes energetic disorder with standard deviation $\sigma = \SI{100}{meV}$. Disorder greatly lowers the potential barrier the charges need to overcome before separating. The free-energy barriers $\Delta G^\ddagger$ are free-energy differences between $r = \SI{1}{nm}$ and $r^\ddagger$, the separation with maximum $\Delta G$.} 
\label{fig:barrier}
\end{figure}

An important consequence of introducing an entropic contribution to the free energy is the appearance of a maximum in $\Delta G(r)$ as shown in Fig.~\ref{fig:barrier}. The position $r^\ddagger$ of this maximum is a more natural definition of the radius at which the electron and the hole separate than the Bjerrum length $r_c$, since the charges will spontaneously move apart past $r^\ddagger$. In the energetically ordered case, if $\Omega(r) \propto r^\gamma$, then $r^\ddagger = r_c/\gamma$. With a fixed hole, $\gamma = 2$ predicts $r^\ddagger=r_c/2\approx \SI{8}{nm}$, whereas allowing the hole to move reduces $r^\ddagger$ further to about $r_c/3\approx \SI{5}{nm}$ (or exactly $r_c/3$ if the initial CT state separation distance is $0$).  Nevertheless, the barrier is still considerable, $\sim 7k_\mathrm{B}T$.

\begin{figure*}[t!]
\centering
\includegraphics[width=0.9\linewidth]{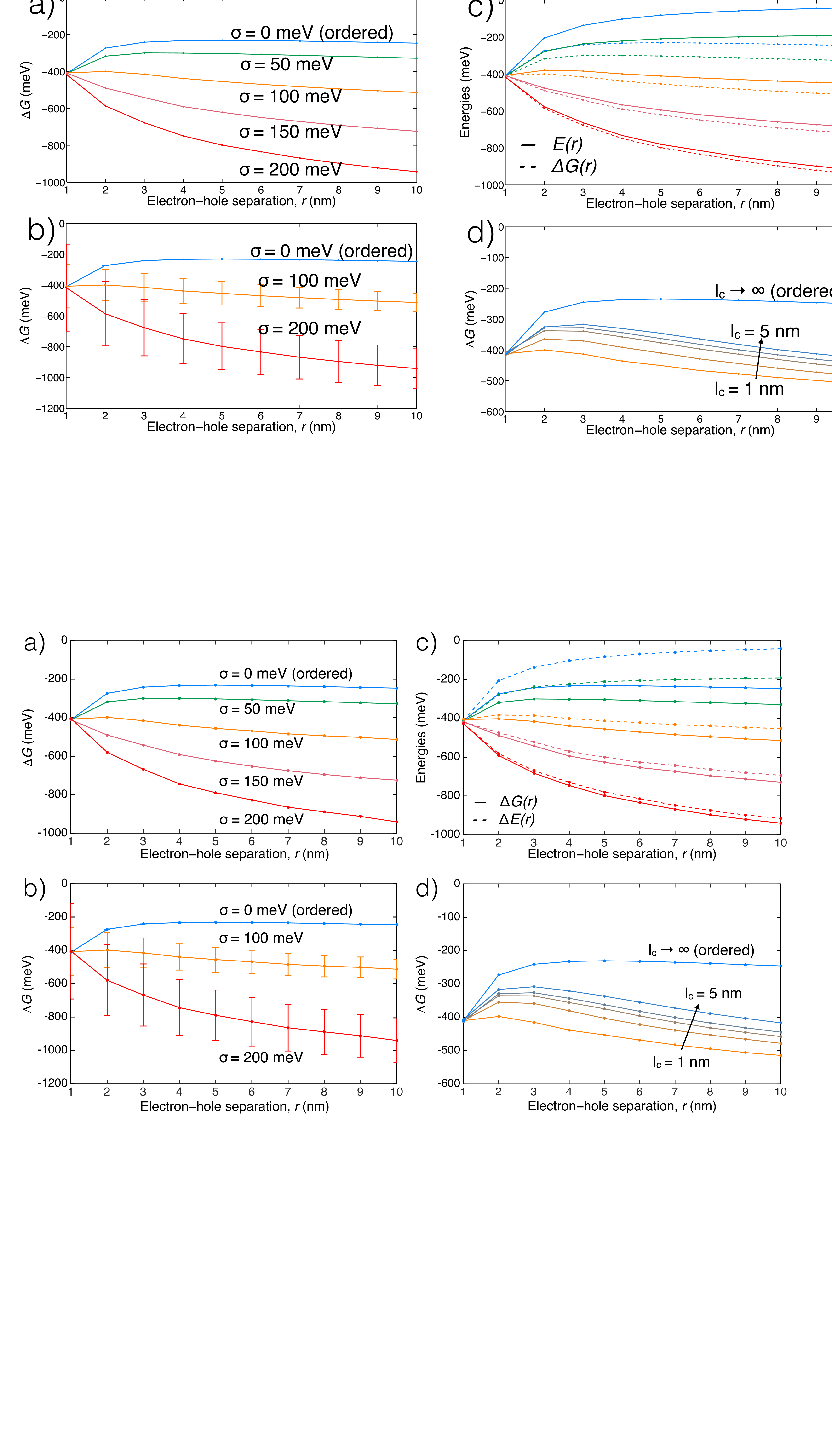}
\caption{The effect of energetic disorder on the free energy of the electron-hole pair.
(a) As the disorder $\sigma$ increases, the potential barrier decreases and, for $\sigma \gtrsim \SI{150}{meV}$, disappears altogether. Each value of $\Delta G$ is the mean of 1000 realisations of the disorder. 
(b) Standard deviations of the distributions of $\Delta G$ (fewer values of $\sigma$ are shown for clarity). The standard deviations of $\Delta G$ are comparable to $\sigma$ indicating that---in particular realisations of the disorder---a low-lying trap state can lower the free energy of its spherical shell substantially below that of its neighbours, preventing the charges from coasting apart down the free-energy slope.
(c) The contributions of entropy and disorder to lowering the charge-separation barrier can be distinguished. The dotted lines show the thermally averaged energies of CT state with a given separation $r$, i.e., the extent to which the barrier is lowered by disorder alone. The solid lines show the further reduction due to entropy, which decreases as disorder increases. The colours of the lines correspond to the energetic disorder as in panel (a).
(d) The free-energy barrier increases with increasing correlation length $l_c$ because correlations effectively smooth the energetic landscape, leading to less disorder. These results are for $\sigma = \SI{100}{meV}$. $l_c = \SI{1}{nm}$ does not produce correlations because it is less than the lattice spacing; therefore, the bottom curve in panel (d) is equal to the $\sigma=\SI{100}{meV}$ curve in panel (a).}  
\label{fig:FreeEnergy} 
\end{figure*}

OSCs are disordered materials and disorder can significantly affect the entropy. We introduce energetic disorder by drawing the site energies $E_i$ from a normal distribution with standard deviation $\sigma$ \cite{Bassler1993,Bassler2015}. In the presence of disorder, electron-hole arrangements with equal separation no longer have the same energy and their thermodynamic properties are most easily calculated in the canonical ensemble. For a system at constant temperature, the free energy (in the absence of $PV$ work, the Helmholtz and Gibbs free energies are equal) is 
\begin{equation}
\Delta G_\mathrm{disordered}(r) = -\langle k_\mathrm{B}T\ln Z(r) \rangle,
\label{eq:G}
\end{equation} 
where $\langle \cdot \rangle$ denotes averaging over the disorder ensemble (we used 1000 realisations for each value of $r$). The partition function is 
\begin{equation}
Z(r) = \sum_{\alpha=1}^{\lfloor\Omega(r)\rfloor} \exp\left(- \frac{U(r) + E^e_\alpha + E^h_\alpha}{k_\mathrm{B} T}\right),
\label{eq:Z}
\end{equation}
where the sum goes over all possible arrangements $\alpha$ of the electron and the hole at a distance $r$ apart. The number of arrangements $\lfloor\Omega(r)\rfloor$ is taken from Eq.~\ref{eq:omega} (and rounded down) to ensure consistency with Eq.~\ref{eq:DeltaG} in the ordered limit $\sigma = 0$. The charge arrangements were determined by selecting $\lfloor\Omega(r)\rfloor$ sites from the hexagonal close-packed lattice closest to the corresponding spherical cap. The energies $E^e_\alpha$ and $E^h_\alpha$ are the disordered site energies of the electron and hole sites in arrangement $\alpha$.

Including disorder significantly reduces the free-energy barrier, as shown in Fig.~\ref{fig:FreeEnergy}a. Disorder allows the charges to separate more easily because they are more likely to find lower-energy sites at greater separations. Although the amount of disorder depends on the material and its preparation, our conclusions hold for disorder close to the typical value of $\sigma = \SI{100}{meV}$~\cite{Gartstein1995,Clarke2010, BasslerBook}. In that case, the height of the energy barrier $\Delta G^{\ddagger} \approx k_\mathrm{B}T$, meaning that the thermal energy could facilitate charge separation (see Fig.~\ref{fig:barrier} for comparison with ordered models). 

Fig.~\ref{fig:FreeEnergy}a depicts the average free-energy landscape due to the averaging in Eq.~\ref{eq:G}. In reality, each realisation of the disorder results in a different landscape, and the corresponding standard deviations in $\Delta G_\mathrm{disordered}(r)$ are shown in Fig.~\ref{fig:FreeEnergy}b. In particular, the presence of disorder means that some free-energy landscapes contain low-lying traps which may hinder charge separation. 

It is possible to separate the contributions of entropy and disorder to the free energy in Eq.~\ref{eq:G}. By itself, the disorder lowers the energy from $U(r)$ to
\begin{equation}
\Delta E_\mathrm{disordered}(r) = \left\langle \frac{1}{Z(r)} \sum_{\alpha=1}^{\lfloor\Omega(r)\rfloor} E_\alpha e^{- E_\alpha/k_\mathrm{B} T} \right\rangle,
\label{eq:energy}
\end{equation} 
where $E_\alpha = U(r) + E^e_\alpha + E^h_\alpha$, while the difference between $\Delta G_\mathrm{disordered}(r)$ and $\Delta E_\mathrm{disordered}(r)$ is the purely entropic contribution, as illustrated in Fig.~\ref{fig:FreeEnergy}c. The entropic contribution is largest in the ordered case, where the occupation of all states at fixed $r$ is equally likely, whereas increasing disorder decreases the entropic component because the charges are most likely to be found in a smaller number of lower-lying sites. Consequently, in very disordered materials the disorder dominates the reduction in free energy, whereas at medium values of $\sigma$ the effects of entropy and disorder are comparable.

Spatial correlations in the energetic disorder were also investigated to determine how local energetic smoothing affects the free energy of charge separation. The correlated site energies $\tilde{E}_j$ were obtained by
\begin{equation}
\tilde{E}_j = N_j^{-1/2} \sum_{i=1}^{N_j} K(r_{ij})E_i,
\end{equation}
where $N_j$ is the number of sites within $l_c$ of site $j$ and $K(r_{ij})$ equals 1 if $r_{ij} \leq l_c$ and 0 otherwise~\cite{Gartstein1995}. The factor $N_j^{-1/2}$ ensures that $\tilde{E}_j$ also have standard deviation $\sigma$. 

The expression for the free energy $\Delta G(r)$ with correlated disorder is the same as in the uncorrelated case, except that the energies $E^{e,h}_\alpha$ in Eq.~\ref{eq:Z} are replaced with their correlated counterparts $\tilde{E}^{e,h}_\alpha$. 

As the disorder correlation length $l_c$ increases, the height of the free-energy barrier also increases. Correlations create a more ordered energetic landscape, decreasing the entropic contribution to the free energy, as seen in Fig.~\ref{fig:FreeEnergy}d. The effect of increasing correlations is roughly equivalent to decreasing the energetic disorder, and so it is sufficient to know the effective disorder $\sigma$ without needing to know the exact extent of spatial correlation.

Overall, including a mobile hole and a disordered energy landscape predicts a lower barrier and a more realistic barrier position compared to previous work. In particular, our prediction of a small $r^\ddagger$ is consistent with the experimental evidence that the charges separate when they are about \SI{4}{nm} apart~\cite{Gelinas2014,Barker2014}. The finding that disorder enhances charge separation is consistent with previous work indicating high charge generation in OSCs with an energetically disordered interface~\cite{Zimmerman2012,Burke2014}. By contrast, in the bulk---once the charges are separated---disorder reduces performance because it increases the likelihood of charge trapping, thus decreasing charge mobility and extraction~\cite{Gartstein1995,Groves2010,Burke2014}. 

Our assumption of localised charge carriers shows that our conclusions hold in the worst-case model with the most strongly bound charges. In reality, both charges are partially delocalised \cite{Bernardo2014, Tamura2013, Barker2014, Bittner:2014ea, Gagorik2015}, which would make it even easier for them to surmount the small barrier \cite{Gagorik2015}. Given that our model with uncorrelated disorder $\sigma = \SI{100}{meV}$ predicts $r^\ddagger = \SI{2}{nm}$, delocalised charges could well start off on the far side of the barrier.

We emphasise that our results concern the thermodynamics of charge separation and not the kinetics and thus cannot be used to calculate precise rates or efficiencies. Indeed, a low or non-existent barrier to charge separation makes the Arrhenius equation inappropriate, since a quasi-equilibrium between the reactants and the transition state will not be established. Ultimately, the efficiency of charge separation depends on the kinetic competition between separation and recombination, and our claim---that the Coulomb binding is not the dominant barrier to charge separation---does not rule out other kinetic obstacles that might make separation slower than recombination. In particular, we do not predict that all disordered OSCs have high efficiencies, since efficient separation requires not only a low barrier, but also a high frequency of attempts to cross it. Indeed, slow charge separation could yield a low efficiency even if there were no thermodynamic barrier at all. However, the insignificance of the Coulomb barrier does mean, for example, that attempts to improve charge yields by reducing the Coulomb attraction---e.g., by increasing $\varepsilon_r$~\cite{Koster2012,Torabi2015}---are, by themselves, unlikely to result in the dramatic improvements.

Similarly, our model does not apply to exciton dissociation in a single phase, where it might be expected to predict high yields that are not observed. The main difficulty with interpreting a localised exciton as a pair of point charges is that their separation would be zero, making the Coulomb barrier infinite. Even if the potential between the charges were modified to account for a finite exciton binding energy, there would be less entropy to be gained from charge separation because, in a single phase, the exciton is not restricted to an interface. In particular, the number of configuration would scales as $r^2$~\cite{Gregg2011}, instead of $r^3$ when an interface is present. More importantly, the recombination lifetimes of CT states are much longer than exciton recombination lifetimes~\cite{Clarke2010, Westenhoff2008, Howard2010}, meaning that CT states have more opportunities for dissociation, giving a higher efficiency. 

An alternative, kinetic perspective on charge separation is afforded by kinetic Monte Carlo simulations, which can also predict efficient charge separation for similar parameter values~\cite{Offermans2005,VanEersel2012,Zimmerman2012,Burke2014}. Analytical treatments based on Onsager theory~\cite{Onsager1934,Onsager1938,Sano1979,Braun1984,Wojcik2009,Hilczer2010,Wojcik2010} are also instructive, even if they are unable to treat disorder. All of these kinetic simulations capture entropy indirectly, since there are more pathways for the charges to move apart than there are for them to move together. Our thermodynamic point of view is a complementary one, with the advantage of a conceptually simple argument that directly refutes the widespread view that charge separation in OSCs involves surmounting a hopelessly large barrier.

In summary, our results show that entropy and disorder can drive charge separation even if the charges are localised and thermally relaxed. For typical energetic disorder $\sigma = \SI{100}{meV}$, the height of the free-energy barrier is about $k_\mathrm{B}T$, which means that charge separation can occur without a large thermodynamic barrier as previously thought. Furthermore, our prediction of a small $r^\ddagger$ is consistent with experimental evidence that shows that $\SI{4}{nm}$ separation is sufficient for generating a charge separated state~\cite{Gelinas2014,Barker2014}. In the future, our estimates of barrier height will be refined by including delocalisation, morphological variation, and energetic gradients caused by external electric fields or molecular aggregation~\cite{Few2014}. However, since we started with the worst-case model of localised, relaxed charges, additional enhancement mechanisms would only strengthen our argument that there is no thermodynamic obstacle to charge separation.

We thank Ardalan Armin, Al\'an Aspuru-Guzik, Katherine Gray, Paul Meredith, and Martin Stolterfoht for helpful discussions. This work was supported by the Australian Research Council through a  Discovery Early Career Researcher Award (DE140100433) and through the Centre of Excellence for Engineered Quantum Systems (CE110001013).

\end{document}